%% file: main.tex
\documentclass[12pt]{article}
\usepackage{graphicx}

\usepackage{color} 

\begin{document}
\newtheorem{defn}{Definition}
\newtheorem{prop}{Proposition}
\newtheorem{lem}{Lemma}
\newtheorem{thm}{Theorem}
\newtheorem{rem}{Remark}
\newtheorem{assm}{Assumption}
\newcommand{\bq}{\begin{equation}} 
\newcommand{\eq}{\end{equation}} 
\title{Ecological  Regression with Partial Identification}
 \author{Wenxin Jiang\footnote{wjiang@northwestern.edu, Institute of Finance (Adjunct), Shandong University, and  Department of Statistics, Northwestern University}, 
 Gary King\footnote{king@harvard.edu,  Institute for Quantitative Social Science, Harvard University},  Allen Schmaltz \footnote{schmaltz@fas.harvard.edu, Institute for Quantitative Social Science, Harvard University}, Martin A. Tanner\footnote{mat132@northwestern.edu, Department of Statistics, Northwestern University}  
  }
\maketitle
\vspace*{-0.2in}
 Abstract: Ecological inference (EI) is the process of learning about individual behavior from aggregate data. We study a partially identified linear contextual effects model for EI and describe how to estimate the district level parameter averaging over many precincts in the presence of the non-identified parameter of the contextual effect. This may be regarded as a first attempt in this venerable literature to limit the scope of the key form of non-identifiability in EI. To study the operating characteristics of our model, we have amassed the largest collection of data with known ground truth ever applied to evaluate solutions to the EI problem. We collect and study 459 datasets from a variety of fields including public health, political science, and sociology. The datasets contain a total of 2,370,854 geographic units (e.g., precincts), with an average of 5,165 geographic units per dataset. Our replication data are publicly available via the Harvard Dataverse (Jiang et al. 2018) and may serve as a useful resource for future researchers.  For all real data sets in our collection that fit our proposed rules, our approach reduces the width of the Duncan and Davis (1953) deterministic bound, on average, by about 45\%, while still capturing the true district level parameter in excess of 97\% of the time.


MSC2010 Classification Codes:   62P25, 62J99 


\section{Introduction}

\input{sections/sec1_introduction.tex}


\section{A linear contextual model}\label{seclinearcontext}

\input{sections/sec2_model.tex}


\section{Bounds on the non-identified $w_1$}\label{secboundparam}

\input{sections/sec3_bounds.tex}


\section{Applying the $w_1$-bound to bound the district level parameter}\label{secboundB}

\input{sections/sec4_applyingBounds.tex}


\section{Asymptotic conservative confidence intervals for $B$}\label{secci}

\input{sections/sec5_confidenceIntervals.tex}


\section{Analytic and Simulated Examples}\label{secex}

\input{sections/sec6_examples.tex}


\section{Real data analyses}\label{secreal}

\input{sections/sec7_dataAnalyses.tex}


\section{Generality and limitation of the current work}\label{secgen}

\input{sections/sec8_limitations.tex}


 \section{Discussion}\label{secdisc}

\input{sections/sec9_discussion.tex}


\section*{Acknowledgements}  The first author is partially supported by a special fund from the Taishan Scholar Construction Project.

\section*{REFERENCES}  
\begin{description}

\item Achen, C. H. and  Shively, W.  P.  (1995). Cross-Level Inference. Chicago: University of Chicago Press.

\item Altman, M., Gill, J., and  McDonald, M. (2004).  A Comparison of the Numerical Properties of EI
 Methods, 383-409. (Edited by King, G., Rosen, O. and Tanner, M.A.,  Cambridge University Press, New York.)

\item Centers for Disease Control and Prevention (CDC), National Center for Health Statistics (2017). Underlying Cause of Death 1999-2016 on CDC WONDER Online Database. Data are from the Multiple Cause of Death Files, 1999-2016, as compiled from data provided by the 57 vital statistics jurisdictions through the Vital Statistics Cooperative Program. Accessed at http://wonder.cdc.gov/ucd-icd10.html (retrieved in 2017).

\item Chambers, R. L. and Steel, D. G., (2001). Simple methods for ecological inference in 2 x 2 tables. J R Stat Soc Ser A 164(Part 1): 175-192.  

\item  Chernozhukov, V., Hong, H., and Tamer, E. (2007). Estimation and confidence regions for parameter sets in econometric models.   Econometrica  75,  1243-1284.

\item Cho, W. K. T., and  Manski, C. F. (2008). Cross level/ecological inference. Oxford handbook of political methodology, 530-569. (Edited by in J. Box-Steffensmeier, H. Brady, and D. Collier,   Oxford University Press, Oxford, UK)

\item Duncan, O. D.,  and Davis, B. (1953); An alternative to ecological correlation. American Sociological Review  18, 665-666.

\item  Goodman, L. (1953); Ecological regression and the behavior of individuals. American Sociological Review 18, 663-664.

\item Imai, K., Lu, Y.,  and  Strauss, A. (2008). ``Bayesian and Likelihood Inference for 2 x 2 Ecological Tables: An Incomplete Data Approach.'' Political Analysis, Vol. 16, No. 1 (Winter), pp. 41-69.

\item Jiang, W., King, G., Schmaltz, A., and Tanner, M. A. (2018). ``Replication Data for: Ecological Regression with Partial Identification''. https://doi.org/10.7910/DVN/8TB7GO. Harvard Dataverse, V1.

\item King, G. (1997). A Solution to the Ecological Inference Problem: Reconstructing Individual Behavior from Aggregate Data. Princeton: Princeton University Press.

\item King, G., Rosen, O. and Tanner, M.A., (2004). Ecological Inference: New Methodological Strategies. Cambridge University Press, New York.

\item Liao, Y. and Jiang, W. (2010). Bayesian analysis in moment inequality models.
{\it The Annals of Statistics} 38, 275-316.

\item Office of the Registrar General \& Census Commissioner, India (2001). Census of India 2001. Accessed at https://data.gov.in (retrieved in 2017).

\item Owen, G. and Grofman, B. (1997).
Estimating the likelihood of fallacious
ecological inference: linear ecological
regression in the presence context effects.
  Political Geography  16,  675-690.

\item  Ruggles, S.,  Genadek, K.,  Goeken, R.,  Grover, J. and  Sobek, M. (2017). Integrated Public Use Microdata Series: Version 7.0 [dataset]. Minneapolis, MN: University of Minnesota, 2017.\\ https://doi.org/10.18128/D010.V7.0. Accessed at https://usa.ipums.org/usa/ (retrieved in 2018).

\item Wakefield, J. (2004). Prior and likelihood choices in the analysis of ecological data. Ecological Inference: New Methodological Strategies, 13-50. (Edited by King, G., Rosen, O. and Tanner, M.A.,  Cambridge University Press, New York.)

\end{description}



\section*{Appendix A: A derivation of the confidence interval $CI_x$ in Proposition \ref{propci} and some comments}

\input{sections/sec_appendix_a.tex}

\newpage

\section*{Appendix B:  Non-emptiness of $CI_0$   when the assumptions hold and the sample size is large enough}

\input{sections/sec_appendix_b.tex}


\end{document}

%% file: sections/sec1_introduction.tex

Ecological Inference (EI) may be described as the problem of making inference on the conditional probability distribution, when only the marginal distributions are known. As an example, voting proportions for Trump $T_i$ are known in each precinct $i$, as well as the proportion of African-Americans $X_i$. However,  due to legal reasons, we do not know the proportion of Trump voters among the black voters   beyond what is described by a bound by Duncan and Davis (1953). However, under some assumptions, marginal information over many precincts with different proportions of African-Americans may be used to estimate the conditional information,   as pointed out by Goodman (1953).  Although we will describe EI in terms of the special context of voting behavior with 2x2 tables (such as how black and white citizens vote for Trump or Clinton), King (1997) points out that EI has a wide variety of applications, such as in policy making, epidemiology, marketing, education, and in many similar problems where  detailed information needs to be mined from data at an  aggregated level.
A later book (King, Rosen and Tanner 2004) has collected contributions from 34 authors on many interesting research problems in EI, which testifies to the importance of this research field. 

One approach for  EI  with 2x2 tables is to utilize the model-free bounds of Duncan and Davis (1953). The Duncan and Davis (hereafter, DD) approach has the advantage of not making any assumptions on the data generation process; however, it generally leads to bounds that are relatively wide.
On the other hand, regression approaches in EI presented in Goodman (1953)   can provide sharp point estimates, which can be very sensitive to the underlying model assumptions. 
These assumptions will be referred to as the standard EI assumptions.
They assume, for example, that white voters in precincts in the United States are equally likely to vote for a particular party's candidate, regardless of the proportion of black citizens in the precincts in which they live.

In this paper, we consider a method that  extends the regression model of Goodman (1953), where, for example,  the race specific voting probability can depend linearly on the race proportion of the precinct, with a nonzero slope parameter representing the linear ``contextual'' effect.
This addresses the most important violation of the standard EI assumptions. 
  On the other hand, it is well known (e.g.,  Owen and Grofman 1997,  Chambers and Steel 2001, Wakefield 2004) that modeling dependence of the race specific voting probability on the race proportion by a linear ``contextual effect'' model  can lead to non-identifiability. 

For this unidentifiable linear contextual model, the current paper exploits an observation made in King (1997, Chapter 9). In particular, that work points out that  
the precinct-level DD bounds themselves carry some information about the contextual effects.   It is also clear that there is still a lot of remaining uncertainty about the precise values of these contextual effects.  This fits very well in the  framework of interval data regression, regressing the  precinct-level DD bounds against the race proportion. Although interval data regression can not fully identify the regression coefficients, it can provide  identification regions or bounds  (see, e.g., Chernozhukov, Hong and Tamer 2007, Liao and Jiang 2010). We will apply this technique  in order to bound the unidentified regression parameter that represents the linear contextual effect.   
      
Our bound differs from the  model-free bounds   such as DD. We have a regression model for the overall behavior and we initially bound the unidentified contextual effect (or the slope parameter). Only after this,  will  we derive  an implied ``regression bound'' for the district level  race specific voting proportion. This implied regression bound can then be intersected with the DD bound to achieve a reduced length.

There are, however, two concerns related to the use of this method. 
First, the new bound is no longer model-free. It depends on the   linear contextual 
effects assumptions. Violations of the assumptions can cause the resulting intersection bound (i.e., the regression bound intersected with the DD bound) to miss the true district voting proportion, or to even be empty.
Second, even if the assumptions hold, the implied regression bound   is only derived in the limit of large $p$ (the number of precincts), and it can still miss the true district level voting proportion by an amount on the order of $1/\sqrt{p}$.

To address the second concern, we increase the implied regression bound by a multiple of the standard errors on both sides (similar to forming a confidence interval), before intersecting with the DD bound. To address the first concern, we select only data sets where the implied regression bound has a nonempty intersection with the DD bound. These two ideas combined together turn out to work very well on hundreds of 2x2 datasets constructed from census and other data sources, where the ground truth is known. For most  of the selected data sets,   the resulting intersection bounds become on average much shorter than the DD bound, yet still contain the true district level proportion.

 In summary, the current paper makes three main contributions: 
\begin{description}
\item 1. We provide bounds for the linear contextual effects in a  model that violates the standard assumptions for EI, which previous works have tended to avoid due to issues of non-identifiability. 
\item 2. We apply the information obtained on the linear contextual effects to improve the Duncan and  Davis (1953) bound, derived more than sixty years ago,  for the district level race specific voting proportion.
\item 3.  The 459 datasets used to examine the operating characteristics of our model are publicly available via Harvard Dataverse (Jiang et al. 2018) and will serve as a useful resource for researchers in EI, as well as in discrete data model. 
In this way, we have amassed the largest collection of data with known ground truth ever applied to evaluate solutions to the EI problem.   
\end{description}
  
 In the following, we first define the linear contextual model in Section \ref{seclinearcontext} and explain why some of the regression coefficients are not identifiable.  We describe how to bound the unidentified regression coefficients in Section \ref{secboundparam}, and we describe how to bound the the district level voting proportion in Section \ref{secboundB}. In Section  \ref{secci} 
we introduce confidence intervals for the bounds to account for finite sample variation. In Sections \ref{secex} and
\ref{secreal} we provide analytic, simulated, and real data examples.
In Section \ref{secgen} we discuss the generality and limitation of the
proposed model. Section
\ref{secdisc} provides further discussion. Some technical details are left to the Appendix.

%
  
 We now describe the linear contextual model and introduce the notation.

%% file: sections/sec2_model.tex
 


We start with the EI ``accounting identity" for precincts $i=1,...,p$:
\begin{equation}\label{obs1}
T_i=X_i\beta_i^b+(1-X_i)\beta_i^w.
\end{equation}
Here, in our running example, $T_i$ is the proportion of voters in precinct $i$ for a candidate of interest, $X_i$ is the proportion of black voters, $\beta^b_i$ is the proportion of voters for a candidate of interest among the black voters, and $\beta^w_i$ is the proportion of voters for a candidate of interest among the non-black voters.

We now  allow  ``contextual effects'', where the race-specific voting behaviors ($\beta^b_i$  and $\beta^w_i$)  can possibly be dependent on the ``context'' (e.g., the black proportion $X_i$).  The following is the only essential assumption that we make in this paper:
 
\begin{assm} (Linear contextual effects.) 
Assume that  $(  \beta^b_i,  \beta^w_i, X_i   )$, for $i=1,...,p$, are independent and identically distributed ($iid$) random vectors that satisfy  
\bq\label{asw} E(\beta_i^w|X_i )=w_0+w_1X_i  \eq and \bq\label{asb} E(\beta_i^b|X_i)=b_0+b_1X_i,\eq
where $w_0,w_1,b_0,b_1$ are four non-random real parameters.  
 \end{assm}

Under these assumptions, $(\beta^b_i,  \beta^w_i, X_i )$  from each precinct is regarded as a vector of random variables sampled from an underlying probability distribution. The conditional expectations  $E(\beta_i^{b,w}|X_i)$ are taken over the conditional distribution of $\beta_i^{b,w}$ given $X_i$, which allows for $\beta_i^{b,w}$ to still be random even after fixing  the values of $X_i$. For example, precincts with similar $X_i$'s (e.g., around 0.5) can still have very different race-specific voting proportions, $\beta_i^b$ or $\beta_i^w$.  

Under these assumptions, regarding $E(\beta_i^{b,w}|X_i)$, the accounting identity (\ref{obs1}) implies that $T_i$ vs $X_i$ satisfies a quadratic regression model: \bq E(T_i|X_i )=w_0+(b_0-w_0+w_1)X_i +(b_1-w_1) X_i^2 .\eq 
We will refer to \bq \label{cdeq} c_1=b_0-w_0+w_1,\; d_1=b_1-w_1,\eq as the coefficients of $X_i$ and $X_i^2$, respectively. It then follows that
\bq \label{teq} E(T_i|X_i )=w_0+c_1X_i +d_1 X_i^2 .\eq 
 The three parameters $(w_0,c_1, d_1 )$ are identifiable (if the $X_i$'s can take three or more distinct values) and can be estimated by (possibly weighted) least squares regression.

The four regression parameters in the linear contextual effects model  are related to the three regression parameters in the quadratic regression of $T_i$ vs $X_i$ via 
\bq (w_0,w_1, b_0,b_1 )=(w_0, w_1, c_1+w_0-w_1 , d_1+w_1),\label{parrel}\eq
which are partially identifiable up to one free parameter: $ (w_0, c_1, d_1)$ are   identified, but $w_1$ is not.

Regarding  the unknown $w_1$, diverging opinions include setting $w_1=\max\{-d_1,0\}$  (Achen and Shiveley 1995; Altman, Gill, and McDonald 2004),
$w_1=0$ (Wakefield 2004, Section 1.2), or $w_1= -d_1/2$ (Wakefield 2004, Section 1.2).   Our approach differs in an important respect. Instead of picking a value for $w_1$, we will derive a prior-insensitive bound for $w_1$ under the current linear contextual effects model, using the expectations of the Duncan-Davis bounds conditional on the $X_i$'s.\footnote{Although we focus on bounding $w_1$ in this paper, we also could have chosen  $b_1=w_1+d_1$ as the non-identified parameter instead. The results would be equivalent due to the accounting identity (\ref{obs1}). However, in that case a composite parameter  $b_0+b_1$ (instead of simply $w_0$) is identifiable, and the notation would be somewhat clumsier.
}  

%% file: sections/sec3_bounds.tex
\subsection{Intuition behind the bounding}\label{boundingIntuition}
  
 Denote the Duncan-Davis bounds for the unobserved $\beta_i^w$ as $L_i\leq\beta_i^w\leq U_i$, where $L_i\equiv\max\{0, (T_i-X_i)/(1-X_i)\}$, $U_i\equiv\min\{1,T_i/(1-X_i)\}$. 
Under the linear contextual model   $E(\beta_i^w|X_i)=w_0+w_1X_i$, the observable Duncan-Davis bounds $L_i\leq\beta_i^w\leq U_i$ form a problem of interval data regression, regressing $[L_i,U_i]$ against $X_i$. It is well known (see, e.g., Chernozhukov, Hong and Tamer 2007, Liao and Jiang 2010) that although interval data regression can not fully identify the regression coefficients, it can provide their identification regions or bounds.
We will use this perspective to derive a bound for the non-identified regression coefficient $w_1$.

 Since the Duncan-Davis bounds for $\beta_i^w$ is $L_i\leq\beta_i^w\leq U_i$,  the corresponding bound in  the conditional expectation is $E(L_i|X_i)\leq E(\beta_i^w|X_i)\leq E(U_i|X_i)$, or
\bq\label{bdline}
E(L_i|X_i)\leq w_0+w_1X_i\leq E(U_i|X_i),
\eq
 where  $L_i=\min\{0, (T_i-X_i)/(1-X_i)\}$, $U_i=\min\{1,T_i/(1-X_i)\}$.
The upper and lower bounds are identifiable from  observable quantities. Forcing  
 this bound in the entire domain of $X_i$ will lead to a bound for $w_1$.

\begin{figure}[htbp]
\includegraphics{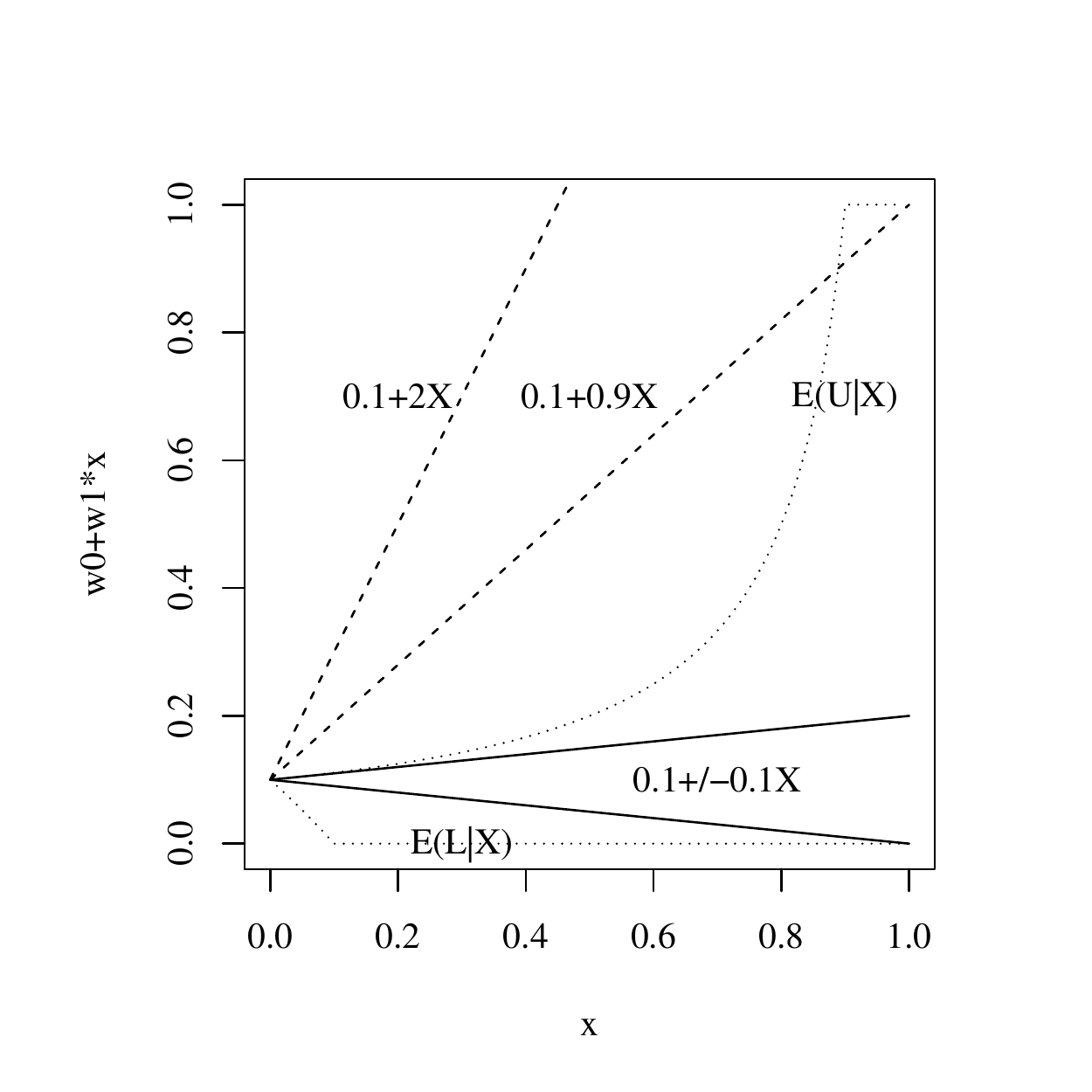}
\caption{Intuition for bounding $w_1$.}\label{fig.int1}
The dotted curves are the expectations of the Duncan Davis bounds.
The solid lines $0.1\pm 0.1x$ are obtained by forcing  linear contextual effects $E(\beta^w_i|X_i=x)=w_0+w_1x$ to lie between the dotted curves. The dashed lines are examples exceeding the expectation of the Duncan-Davis upper bound (see Section \ref{boundingIntuition}).

 \end{figure}

Consider a very simple example where $T_i
\equiv 0.1$ for all $i$.
In Figure \ref{fig.int1}, we illustrate the intuition of how to bound the slope parameter $w_1$ in the linear contextual model $E(\beta^w_i|X_i=x)=w_0+w_1 x$ for all $x\in(0,1)$. The intercept parameter is identifiable as $w_0=E(T_i|X_i=0) = 0.1$.  The   slope parameter $w_1$ is non-identified, but only partially so. There are hidden constraints: if the line $ w_0+w_1 x=0.1+2x$, then the probability  $E(\beta^w_i|X_i=x)=w_0+w_1 x$ can exceed $1$, so $w_1$ cannot be as high as $2$ ($w_1\leq 2$). Even if we choose  $ w_0+w_1 x=0.1+0.9x$ so that $E(\beta^w|X_i=x)=w_0+w_1 x$ falls between $[0,1]$ for all $x\in(0,1)$,  the line $0.1+0.9x$ can still exceed a large portion of the dotted curve of the expectation of the Duncan-Davis upper bound $E(U_i|X_i=x)$. As such, $w_1$ also can not exceed $0.9$ ($w_1\leq 0.9$). In fact, to force $w_0+w_1 x$ to fall between the dotted curves $[E(L_i|X_i=x),E(U_i|X_i=x)]$ for all $x\in (0,1)$, we need to have $w_1\in[-0.1,0.1]$, lying in a very narrow interval in this example. 

Intuitively, this is  how we exploit the expectation of the  DD bounds (\ref{bdline}) to bound the non-identified contextual effect parameter $w_1$.
More formally, we have the following theoretical results.

\subsection{Theory}
The following proposition provides a necessary and sufficient condition for this bound in terms of the only non-identified parameter $w_1$.
\begin{prop} \label{prop1} Assume a linear contextual effect $E(\beta_i^w|X_i)=w_0+w_1X_i$  for all $X_i\in A$ where $A\subset (0,1)$.
Then $$E(L_i|X_i)\leq E(\beta_i^w|X_i)\leq E(U_i|X_i),$$ for all $X_i\in A$, if and only if the non-identifiable parameter $w_1$ satisfies  $$\sup_{X_i\in A} [(E(L_i|X_i)-w_0)/X_i]\leq w_1\leq \inf_{X_i\in A} [(E(U_i|X_i)-w_0)/X_i].$$
 \end{prop}
 
   Proof: If $\sup_{X\in A} [(E(L|X)-w_0)/X]\leq w_1\leq \inf_{X\in A} [(E(U|X)-w_0)/X]$ holds, then for all $X\in A$,  
 $ [(E(L|X)-w_0)/X]\leq w_1\leq   [(E(U|X)-w_0)/X]$.  
 This implies   $E(L|X)\leq   w_0+w_1 X\leq   E(U|X)$ for all $X\in A\subset  (0,1)$.
 
 For the converse:  $E(L|X)\leq   w_0+w_1 X\leq   E(U|X)$ for all $X\in A\subset  (0,1)$
 implies   
 $ [(E(L|X)-w_0)/X]\leq w_1\leq   [(E(U|X)-w_0)/X]$ holds for all $X\in A$. Now we take
 $\inf_{X\in A}$ for both sides of $ w_1\leq   [(E(U|X)-w_0)/X]$,  and take $ \sup_{X\in A}$
  for both sides of $[E(L|X)-w_0)/X]\leq w_1$. Q.E.D.\\

  The above proposition then gives the tightest bound possible on $w_1$. The upper bound and the lower bound are both constructed out of identifiable quantities.  The functions $E(L_i|X_i)$ and $E(U_i|X_i)$ may be estimated by lowess smoothing.
  If for some reason we would like to avoid such nonparametric estimation (e.g., it may not perform   well at boundary values of $X_i$), we can relax the bounds somewhat and incorporate results from a parametric regression  $E(T_i|X_i)=w_0+c_1X_i+d_1X_i^2$. 
  \begin{prop} \label{prop2} For all $X_i\in [l,u]\subset (0,1)$ where $l<u$, assume linear contextual effect $E(\beta_i^w|X_i)=w_0+w_1X_i$ AND a quadratic regression $E[T_i|X_i]=w_0+c_1X_i+d_1X_i^2$.
 Then we have  $$wl\leq w_1\leq wu,$$ where $wl=\max_{x \in \{l,u\}} \max\{-w_0/x, (w_0+c_1+d_1-1 )/(1-x)-d_1\}$ and $wu=\min_{x \in \{l,u\}} \min\{(1-w_0)/x, (w_0+c_1+d_1 )/(1-x)-d_1\}$.  
 \end{prop} 
Proof:   For the bounds in Proposition \ref{prop1},
 note that $E(U_i|X_i)=E[\min\{1,T_i/(1-X_i)\}|X_i]\leq  \min\{1, E[T_i|X_i]/(1-X_i)\} $
  due to Jensen's inequality, and similarly  $E(L_i|X_i)\geq \max\{0, (E[T_i|X_i]-X_i)/(1-X_i)\}$. Now apply a quadratic regression $E[T_i|X_i]=w_0+c_1X_i+d_1X_i^2$.
 Then from Proposition \ref{prop1} we have 
 $$   
\sup_{X_i\in A} \max\{-w_0/X_i, (w_0+c_1-1+d_1X_i)/(1-X_i)\}$$
$$  \leq  w_1\leq  \inf_{X_i\in A} \min\{(1-w_0)/X_i, (w_0+c_1+d_1X_i)/(1-X_i)\}.
 $$
Simplifying these bounds for $A=[l,u]$ with the boundary points leads to the proof.
Q.E.D.

To use Proposition \ref{prop2}, we need to supply the interval $[l,u]$ where we believe the assumptions hold. One could simply use the data range $l=\min X_i$ and $u=\max X_i$ of the data set. However, there may be reasons to either reduce this range (e.g., if there are outliers) or increase this range (if there is a belief that the pattern could be reliably extrapolated to some extent beyond the data range).
If we attempt to check the assumptions when there is no knowledge regarding the ground truth $\beta_i^w$,  we could still use the $(T_i,X_i)$ data to fit a quadratic curve on (0,1), and superimpose it on the scatterplot, using it to rule out unreasonable choices of a range $[l,u]$.  For example, when quadratic regression is based on a scatterplot  limited in a small domain of $X_i\in[0.5,0.6]$, and extrapolating the fitted quadratic curve to $x\in [0.1,0.9]$ leads to $E(T|X=x)=w_0+c_1x+d_1x^2$ breaking the ``ceiling" of 1 or the ``floor" of 0,  then it is obvious that the range $[l,u]= [0.1,0.9]$ is too wide.
 
On the other hand, the bigger the set $A=[l,u]$ is for $X_i$, the tighter the bounds will be in these propositions.
Suppose we consider a special case $A\rightarrow (0,1)$. In other words,  we assume   that the previous quadratic regression model   holds for all $X_i$ in the whole range of $(0,1)$.  Then  relaxing the bounds of Proposition \ref{prop2} and taking $l\rightarrow 0$, $u\rightarrow 1$,  we immediately  have:
\begin{prop}\label{prop3}For all $X_i\in  (0,1)$, assume linear contextual effect $E(\beta_i^w|X_i)=w_0+w_1X_i$ AND a quadratic regression $E[T_i|X_i]=w_0+c_1X_i+d_1X_i^2$.
 Then we have  $$wl\leq w_1\leq wu,$$ where $wl=\max\{-w_0, c_1+w_0-1\}$ and $wu=\min\{1-w_0, c_1+w_0 \}$. 
 \end{prop}

The above are the bounds for the non-identified contextual effect parameter $w_1$.
 In practice, we are interested in predicting the   district level  race-specific vote. In the next section, bounds for such quantities are derived from any bound on $w_1$. 


%% file: sections/sec4_applyingBounds.tex
 \subsection{Estimating the district level parameter $B$ by a point estimate $B(\lambda,w_1,\theta)$   given $w_1$, $\lambda$, and $\theta$.}
We first analyze the precinct level parameter $\beta_i^b$. 
Denote residuals as $e_i^b=\beta_i^b-E(\beta_i^b|X_i )$, $e_i^w=\beta_i^w-E(\beta_i^w|X_i )$, and $e_i^T=T_i-E(T_i|X_i )=e_i^bX_i+e_i^w(1-X_i)$.
Note that for any real $\lambda$ possibly dependent on $i$,

$$\beta_i^b=E(\beta_i^b|X_i )+\lambda (T_i-E(T_i|X_i )) +e_i^b-\lambda e_i^T$$
$$\stackrel{by\ (\ref{parrel})}{=}w_0+(c_1-w_1)+(w_1+d_1)X_i  + \lambda(T_i-w_0-c_1X_i -d_1X_i^2 )
+e_i^b-\lambda e_i^T$$
$$=[w_0+c_1+d_1X_i  ]+ \lambda(T_i-w_0-c_1X_i- d_1X_i^2 )+w_1(X_i-1)
+(e_i^b-\lambda e_i^T)$$
\bq \equiv b_i(\lambda,w_1,\theta) + (e_i^b-\lambda e_i^T ).\label{bexp}\eq
Here $\theta\equiv (w_0,c_1 ,d_1 )^T$.

The {\em district level parameter} is given as

\bq B\equiv \sum_{i=1}^p N_iX_i \beta_i^b/\sum_{i=1}^p N_iX_i \label{defnB}\eq
$$
=\frac{\sum_{i=1}^p N_iX_ib_i(\lambda,w_1,\theta)}{\sum_{i=1}^p N_iX_i} +\frac{\sum_{i=1}^p N_iX_i(e_i^b-\lambda e_i^T )}{\sum_{i=1}^p N_iX_i}
.$$
Here, $N_i$ denotes the size of the $i$th precinct, which will be incorporated as another random variable in the $iid$ framework, in order to allow the precincts to have different sizes.
It is possible to extend our work to include additional covariates and treat $N_i$ (or its suitable transformation) as a covariate. Here, we simply modify our assumption on the linear contextual effects  so that (\ref{asw}) and (\ref{asb}) hold conditional on both $X_i$ and $N_i$ on the entire support of these random variables.

Due to the Law of Large Numbers, for large $p$, we can ignore the second term of the expansion of (\ref{defnB}) with the mean zero residuals $(e_i^b-\lambda e_i^T )$,  when estimating $B$. We can then form a {\em point estimate} of $B$ using the first term:

$$B(\lambda,w_1,\theta)\equiv \frac{\sum_{i=1}^p N_iX_ib_i(\lambda,w_1,\theta)}{\sum_{i=1}^p N_iX_i}$$
\begin{equation}\label{Bformula}=\frac{\sum_{i=1}^p N_iX_ib_i(\lambda,0,\theta)}{\sum_{i=1}^p N_iX_i}- w_1\frac{\sum_{i=1}^p N_iX_i(1-X_i)}{\sum_{i=1}^p N_iX_i},\end{equation} 
where
\begin{equation}\label{bformula} b_i(\lambda,w_1,\theta)\equiv [w_0+c_1+d_1X_i  ]+ \lambda(T_i-w_0-c_1X_i- d_1X_i^2 )+w_1(X_i-1).\end{equation}

\subsection{The district level point estimate  $B(\lambda,w_1,\theta)$ is insensitive to $\lambda$ and sensitive to $w_1$.} 

Note that $$B(\lambda,w_1,\theta)-B(0,w_1,\theta)=  \frac{\sum_{i=1}^p N_iX_i\lambda e^T_i}{\sum_{i=1}^p N_iX_i}.$$ This is  an average of mean-0 residuals, which will be approximately 0 due to the Law of Large Numbers. Therefore we know that  $B(\lambda,w_1,\theta)$ varies little with $\lambda$ for large $p$.

Different choices of $\lambda$ will therefore make very little difference for large values of $p$. The choice of $\lambda$ may still influence  the  district level parameter  at the order of $O_p(1/\sqrt{p})$. A possible approach to optimizing $\lambda$ at this finer level will be discussed in Remark \ref{remlambda}, but it will be mainly left as  possible future work.  
  In this paper, we   choose $\lambda=1$ for illustration.  

 In contrast to its insensitive dependence on $\lambda$,  the point estimate $B(\lambda,w_1,\theta)$  will vary greatly with $w_1$ due to (\ref{Bformula}), at a level that does NOT disappear for large $p$. The sensitivity on $w_1$ can be measured by 
\bq
\frac{\partial B(\lambda,w_1,\theta)}{\partial  w_1}=-r\equiv -\frac{\sum_{i=1}^pN_iX_i(1-X_i)} {\sum_{i=1}^pN_iX_i},
\label{eqsensitive}
\eq 
 which is typically   nonzero (unless $X_i\in\{0,1\}$ for all nonempty precincts).  This term quantifying the sensitivity on $w_1$  does not converge to 0 even for large $p$. Therefore the bounds we derived earlier for $w_1$ will be very useful here for limiting the scope of the influence by $w_1$.  

Now for  any possible value of the partially identified $w_1$, the district level parameter  $B =\sum_iN_iX_i\beta_i^b/\sum_iN_iX_i$ is estimated by the point estimator $B(\lambda,w_1,\theta)$ following (\ref{Bformula}).
We will now use the bounds on $w_1$ to bound this district level parameter  estimate $B(\lambda,w_1,\theta)$, and estimate its $\theta$ parameter by regression. 

 \subsection{Bounding the  district level point estimate $B(\lambda,w_1,\theta)$  by $[\hat Bl,\hat Bu]$, for unknown $w_1$, with $\theta$ estimated by $\hat\theta$.}

 Due to Propositions \ref{prop2} or \ref{prop3}, we know that
$w_1\in[wl,wu]$,  where  $wu=wu(\theta)$ and $wl=wl(\theta)$ depend on $\theta$.  Then \bq B(\lambda,w_1,\theta)  \in[Bl, Bu] \equiv [B(\lambda,wu(\theta),\theta), B(\lambda,wl(\theta),\theta)].\label{intbd}\eq

The parameters
$\theta=(w_0, c_1,  d_1)^T $  can be estimated from a least squares regression
 \bq\label{regr}
\hat\theta=(\hat w_0, \hat c_1,  \hat d_1 )^T\leftarrow \min_{w_0, c_1,  d_1 } \frac{ \sum_{i=1}^p \rho_i [T_i- (w_0+c_1X_i +d_1 X_i^2 )]^2}{ \sum_{i=1}^p \rho_i},
\eq 
possibly weighted by some weight $\rho_i$.

Replacing $\theta$ in (\ref{intbd}) by $\hat\theta$, we obtain the estimated bounds for the  district parameter  $B$.   Since this is implied from a regression model of linear contextual effects, one may  call this a {\em ``regression  bound"}. In summary,   this will be our proposed  interval estimate for $B$:

\begin{defn}(Regression  Bound.)
A Regression Bound for  the district parameter  $B =\sum_iN_iX_i\beta_i^b/\sum_iN_iX_i$ is of the form 
\begin{equation}\label{intest} [\hat Bl, \hat Bu]\equiv [B(\lambda,wu(\hat\theta),\hat\theta), B(\lambda,wl(\hat\theta),\hat\theta)],\end{equation}
where the functional form of the point estimate $B(\lambda,w_1,\theta)$ follows (\ref{Bformula}),  $wu=wu(\theta)$ and $wl=wl(\theta)$  are  the bounds of the $w_1$ parameter according to Proposition \ref{prop2} or Proposition \ref{prop3}, and  $\hat\theta$ estimates the regression coefficients $\theta$  from (\ref{regr}).
\end{defn}

%% file: sections/sec5_confidenceIntervals.tex
The previous regression bound $[\hat Bl,\hat Bu]$ for $B$ does not take into account sampling variation. It assumes, for example, that the quadratic regression coefficients $\hat w_0, \hat c_1, \hat d_1$ are the true coefficients,  while in reality they are estimated from $p$ precincts and are subject to sampling error.   Due to sampling error, it may be possible that  according to the sample estimates,    $B\not\in [\hat Bl, \hat Bu  ]$, even if the model assumptions for linear contextual effects are valid, when  we should automatically have  $B\in[\hat Bl,\hat Bu]$  in the large $p$ limit. (See Appendix B.)  To solve this problem, we will provide   asymptotic conservative confidence intervals for  $B$  in this section, where $\hat Bl$ will be reduced (and $\hat Bu$ will be increased) by a typical size of the sampling variation. 

Since  $[\hat Bl, \hat Bu]\equiv [B(\lambda,wu(\hat\theta),\hat\theta), B(\lambda,wl(\hat\theta),\hat\theta)]$ depends on the functional forms of $wl(\cdot)$ and $wu(\cdot)$, 
we first need to analyze in detail these functional forms.

In Propositions \ref{prop2} and \ref{prop3}, the bounds $wl$ and $wu$ are functions of the quadratic regression coefficients $\theta=(w_0,c_1,d_1)^T$.  The lower bounds can be expressed in the form \bq \label{eqwl} wl(\theta) =\max_{j=1}^J\{gl^0_j+gl^T_j\theta\},\eq
and the upper bounds can be expressed in the form \bq\label{eqwu}  wu(\theta)=\min_{j=1}^J\{gu^0_j+gu^T_j\theta\}.\eq \\

For Propositions \ref{prop2}, $J=4$, 

$gl^0_1=0$, $gl_1^T=(-1/l,0,0)$,  $gl^0_2=-1/(1-l)$, $gl_2^T=(1/(1-l),1/(1-l), 1/(1-l)-1)$,

$gl^0_3=0$, $gl_3^T=(-1/u,0,0)$,  $gl^0_4=-1/(1-u)$, $gl_4^T=(1/(1-u),1/(1-u), 1/(1-u)-1)$,

 $gu^0_1=1/l$, $gu_1^T=(-1/l,0,0)$,  $gu^0_2=0$, $gu_2^T=gl_2^T$,

 $gu^0_3=1/u$, $gu_3^T=(-1/u,0,0)$,  $gu^0_4=0$, $gu_4^T=gl_4^T$.\\
 
For Proposition \ref{prop3}, $J=2$,

$gl^0_1=0$, $gl_1^T=(-1,0,0)$,  $gl^0_2=-1$, $gl_2^T=(1,1,0)$,

 $gu^0_1=1$, $gu_1^T=(-1,0,0)$,  $gu^0_2=0$, $gu_2^T=(1,1,0)$.\\

Using this notation, we have the following result:
 
\begin{prop}\label{propci} Let $ B= \frac{\sum_{i=1}^p N_iX_i \beta_i^b}{ \sum_{i=1}^p N_iX_i} $ be the district parameter of  voting proportion  for a candidate of interest among all the black people in a district with $p$ precincts.  Let  $DD$ be the Duncan and Davis (1953) bound for $B$, following \begin{equation}\label{ddforb}
DD=\left[  \frac{\sum_{i=1}^p N_i \max\{ 0, T_i-(1-X_i)\} }{ \sum_{i=1}^p N_iX_i},       \frac{\sum_{i=1}^p N_i \min\{ T_i, X_i\} }{ \sum_{i=1}^p N_iX_i}     \right].\end{equation}  For any choice of $\lambda\in[0,1]$,  as $p\rightarrow\infty$, an asymptotic conservative  confidence interval  for $B$  of the form:
 \begin{equation}
CI_x\equiv  [ \hat BL -xSL,  \hat BU+xSU ]\cap DD,
\end{equation}
has  asymptotic coverage probability at least $ \Phi( x)$. 

Here we use the following system of notation:

$x>0$,

$\hat\theta^T=(\hat w_0,\hat c_1,\hat d_1)$  which is estimated by quadratic
 regression (\ref{regr}), which has robust sandwich asymptotic variance matrix $V$,\footnote{See, e.g., https://www.stata.com/manuals/p\_robust.pdf}



$\lambda \in[0,1]$,

$r\equiv \frac{\sum_iN_iX_i (1-X_i)}{\sum_iN_iX_i}$, 

$h_0\equiv \frac{\sum_iN_iX_i  \lambda T_i }{\sum_iN_iX_i}$,

$h^T\equiv\frac{\sum_iN_iX_i (1-\lambda,  1-\lambda X_i,   X_i-\lambda X_i^2)}{\sum_iN_iX_i}$,

  $S_1= \sqrt{  \sum_i\left(\frac{N_iX_i[(1+\lambda)/2-\lambda X_i]    }{ \sum_iN_iX_i}\right)^2 }$,

$\hat BL =\max_{j=1}^J\{\hat BL_j\}$,

$\hat BU =\min_{j=1}^J\{\hat BU_j\}$.

For $j=1,...,J$, the $gl_j$'s and $gu_j$'s are defined after (\ref{eqwl}) and (\ref{eqwu}),

$\hat BL_{j}= h_0-rgu^0_{j}+(h- rgu_{j})^T\hat \theta$,

$\hat BU_{j}= h_0-rgl^0_{j}+(h- rgl_{j})^T\hat \theta$,

 $SL_{j}\equiv S_1+\sqrt{(h-rgu_{j})^TV(h-rgu_{j})}$,

 $SU_{j}\equiv S_1+\sqrt{(h-rgl_{j})^TV(h-rgl_{j})}$,

$SL=SL_{\hat j}$ where $\hat j\equiv \arg \max_{j=1}^J\{\hat BL_j\}$,

$SU=SU_{\tilde j}$ where $\tilde  j\equiv \arg \min_{j=1}^J\{\hat BU_j\}$. \\

For this result to hold, we assume that the linear contextual model holds conditional on both $N_i$ and $X_i$ on the entire support of these random variables, and also for all $X_i$ in a range specified in either Proposition \ref{prop2} or Proposition \ref{prop3}. We assume that the robust variance $V$ is of order $O_{\rm p}(1/p)$.   In addition, we assume   the following ``tie-breaking'' conditions:

\begin{description}

\item (i) Assume that  $N_iX_i(1-X_i)$ is not almost surely 0.
 
\item (ii) Assume that the  minimizing entry of $wu= \min_{j=1}^J\{gu^0_j+gu_j^T\theta \}$ is  unique and not tied with the other entries, and similarly   the  maximizing entry of $wl= \max_{j=1}^J\{gl^0_j+gl_j^T\theta \}$ is  unique and not tied with the other entries.

\item (iii) Assume that $wu(\theta)\neq wl(\theta)$.  

\end{description}
\end{prop}

A derivation of this confidence interval $CI_x$  in Proposition \ref{propci} is included in  Appendix A. 

\begin{rem}The tie breaking conditions can be checked by examining the data at hand. The condition on $N_i$ and $X_i$ is satisfied if $N_i$ is not almost surely 0 and if $X_i$ does not almost surely take a boundary value (0 or 1) for nonempty precincts with $N_i>0$.
The conditions on $\theta$ will hold for almost all true parameters  (except on a set with Lebesgue measure 0, where some of the $2J$ points 
$\{gu^0_j+gu_j^T\theta,  gl^0_j+gl_j^T\theta, j=1,...,J\}$ are exactly tied). In the Bayesian sense when $\theta$ is regarded as a vector of continuous random variables, these conditions hold with probability one, since any ties would force $\theta$ to lie on a lower dimensional manifold which has zero Lebesgue measure.
\end{rem}

\begin{rem}
Instead of the analytic method described here, one may consider using the bootstrap to estimate the standard deviation (sd) of the bound estimate $\hat BL $ (and similarly for $\hat BU $), and replace the $SL$ in the formula of  $CI_x$ by $ sd_{boot}(\hat BL) $. However, we suspect that this bootstrap method  would not be theoretically valid here. The reason is that we are not interested in how much $\hat BL$ varies from its own non-stochastic large sample limit, i.e., the typical size of $\hat BL - lim_{p\rightarrow \infty}\hat BL$. We are really interested in the typical size of $\hat BL -B$ instead. However,  the district level parameter $ B= \frac{\sum_{i=1}^p N_iX_i \beta_i^b}{ \sum_{i=1}^p N_iX_i} $ is an non-identified stochastic quantity, and   its sampling variations   would be ignored by bootstrapping $\hat BL$ alone.  Nevertheless, in practice, the bootstrap method may still work well  heuristically for describing the sampling variation.
\end{rem}


\begin{rem}\label{remlambda}
The proposed confidence interval 
$[\max_j\{\hat BL_j\}-xSL, \min_j\{\hat BU_j\}+xSU\}]$  has width $ \min_j\{\hat BU_j\}-\max_j\{\hat BL_j\}+x(SL+SU) =r(wu(\hat\theta)-wl(\hat \theta))+ x(SL+SU)$, where 
only $S=SU+SL$ depends on $\lambda$.
It is possible to define the best $\lambda\in[0,1]$ by minimizing  $S$, but we will leave this for future work.  


\end{rem}

In the numerical examples below, we currently simply choose $\lambda=1$ for illustration. 

%% file: sections/sec6_examples.tex
We will  compare the proposed bound $CI_x$ to the   Duncan and Davis (1953) bound  $DD$,
as defined in Proposition \ref{propci}. For any interval $A$, we will use $|A|$ to denote its length. 
We will use $x=0$ and $x=1$ for illustration.

To measure the success of the proposed method, we examine:
\begin{description}
\item 1. whether the new interval estimate contains the true district parameter: $B  \in CI_x$.

\item 2. how narrow the new interval estimate is compared to the DD bound:  the width ratio $WR_x\equiv |CI_x|/|DD|$.
\end{description} 

In the examples below, we assume $X\sim Unif[0,1]$, and $N_i$ is constant for all $i$, unless otherwise stated.  
\begin{description}
\item Example 1: $\beta_i^b=T+\tau(1-X_i)\in[0,1]$, and $\beta_i^w=T-\tau X_i \in[0,1]$, where probability constraints entail $\tau\in\pm\min(T,1-T)$ and $T\in(0,1)$.  Then the plot $T_i$ against $X_i$ is a flat $T_i=T$. Here one can show by Proposition \ref{prop3} that $[wl,wu]=\pm \min(T,1-T)$.
In this case, in the limit of large precincts and large number of precincts (large $N_i$ and $p$),   it can be shown analytically that the true parameter $B \approx E\beta_i^b = T+\tau/3 \in CI_0 \approx T\pm (1/3)\min(T,1-T)
\subset DD \approx [T^2 , 2T-T^2 ]$. Also, $WR_0\equiv |CI_0|/|DD| \approx 1/[3\max(T,1-T) ] \in(1/3,2/3)$.
In summary, the proposed bound tightens the DD bound while still containing the true parameter.
 
\item Example 2: $\beta_i^b=\tau(1-X_i)$, $\beta_i^w=1-\tau X_i$, where $\tau\in [0,1]$. Then the plot $T_i$ against $X_i$ is   $T_i=1-X_i$. Here one can show by Proposition \ref{prop3} that $[wl,wu]=[-1,0]$. 
In this case, in the limit of large precincts and large number of precincts (large $N_i$ and $p$),   it can be shown analytically that $B  \approx E\beta_i^b =\tau/3 \in CI_0 \approx [0,1/3]$, $DD  \approx  [0,1/2]$. Also, $WR_0\equiv |CI_0|/|DD| \approx   2/3$.
In summary, the proposed bound tightens the DD bound while still containing the true parameter.

\item Example 3: $\beta_i^b=0$, $\beta_i^w=1-X_i$. Then the plot $T_i$ against $X_i$ is   $T_i=(1-X_i)^2$.  Here one can show by Proposition \ref{prop3} that $[wl,wu]=[-1,-1]$, so $w_1$ is identified.
In this case, in the limit of large precincts and large number of precincts (large $N_i$ and $p$),   it can be shown analytically that $B=0 $,  $CI_0 \approx [0,0]$,   $DD \approx  [0, 2E\min(X,(1-X)^2)]\approx [0,0.3032767]$. Also, $WR_0\equiv |CI_0|/|DD| \approx 0$. 
 
We now generate $p=1000$ precincts all with population $N_i=150$ for this example. 
For sample estimates based on this finite data set, we
obtain true $B=0$,  $DD =[0,0.301843]$.

We apply Proposition \ref{prop2} for this example with $[l,u]=[\min(X_i),\max(X_i)]=[ 0.001473298,  0.9988792
]$.

  We obtain $\hat Bl=\mbox{ 2.269362e-05} $ and $\hat Bu=\mbox{0.0003054308} $ which are very close to $B=0$, but $CI_0=[\hat Bl, \hat Bu]$ excludes the true $B$ due to sampling variation.
On the other hand, the proposed interval estimate narrowly misses the true $B$ due to sampling variation. 
The  confidence interval $CI_x$ for $x=1 $ is
$[  -0.01146876,  0.01181847] \cap DD =[0,0.01181847]$, which does contain the true $B$ now and is still very narrow.
(Here, intersection with the Duncan Davis bound improves the lower bound to be 0.)
In summary, the regression bound $CI_0$   can miss the true parameter due to sampling variation.
However, after expanding the bound to account for the sampling variation, $CI_1$ does contain the true parameter $B$ and is still much narrower than the DD bound.

\item Example 4: Consider  $p=1000$ precincts all with population $N_i=150$. We let $X_i\sim Unif[0,0.95]$, 
$\beta_i^b\approx (N_iX_i)^{-1} Bin(N_iX_i, 1/(1+\exp(-b0-b1*X_i- (1-X_i)*\epsilon^b_i))$, 
$\beta_i^w \approx (N_i(1-X_i))^{-1} Bin(N_i(1-X_i),1/(1+\exp(-w0-w1*X_i- (1-X_i)\epsilon^w_i))$ (the approximation $\approx$ here involves operations such as rounding $N_iX_i$ and adding 1 to avoid zero or fractional counts), where $\epsilon^{b,w}_i$'s are $iid$ $N(0,s^2)$, $s=0.5$, $b0= 2.197225$, $b1=-1.791759$, $w0= 2.197225$, $w1=0$,
  $T_i\approx \beta^b_iX_i+\beta_i^w (1-X_i)$ (the approximation $\approx$ here involves operations such as replacing $X_i$ by a rounded version of $N_iX_i$ divided by $N_i$). The resulting $T_i$ vs $X_i$ scatterplot is given by  Figure \ref{fig.ex4}. 
\begin{figure}[htbp]
\includegraphics{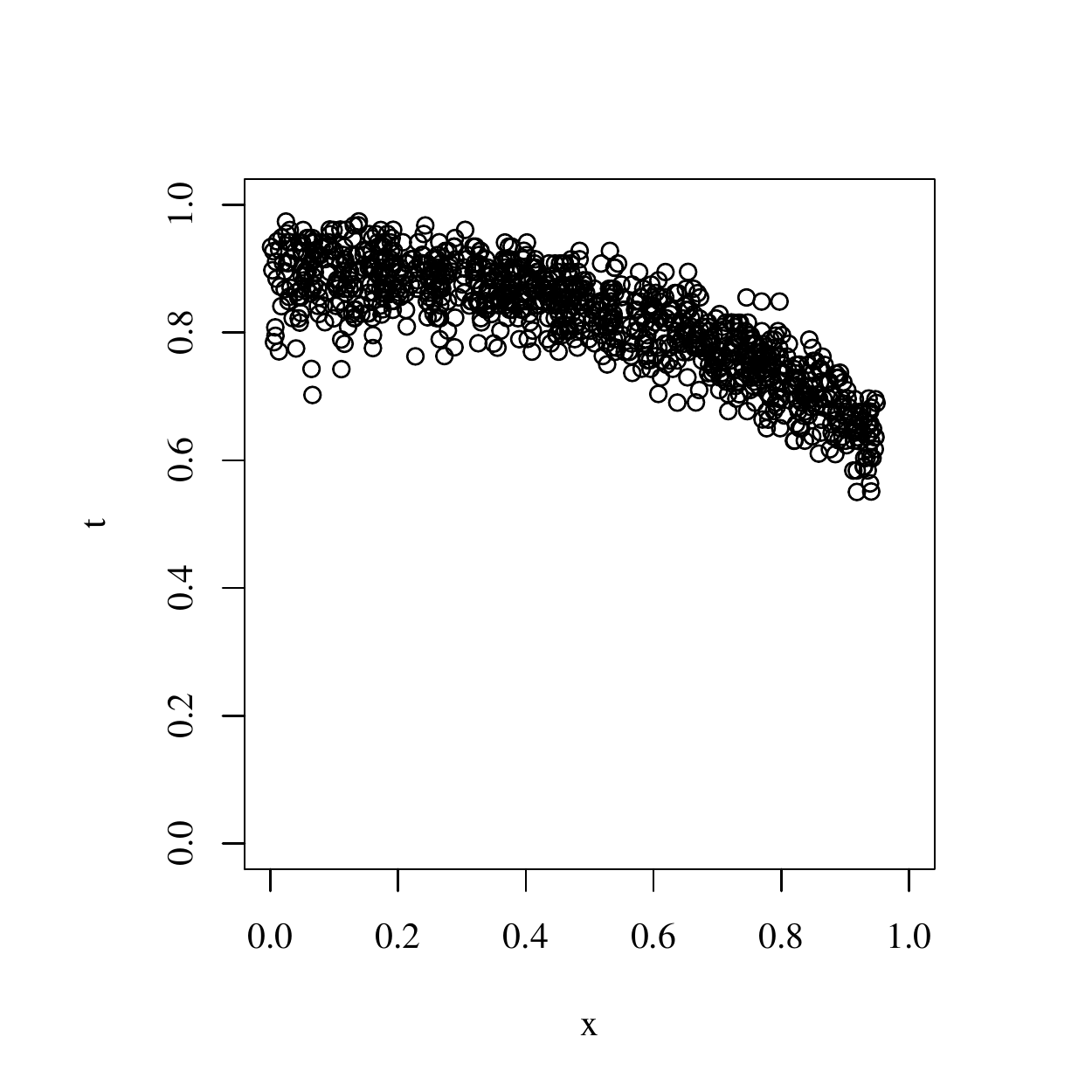}
\caption{T vs X scatterplot for Example 4.}\label{fig.ex4}
 \end{figure} 
%
%

We apply Proposition \ref{prop2} for this example with $[l,u]=[\min(X_i),\max(X_i)]=[0.001399633,  0.9489353]$.

In this case it can be shown that  
 $B= 0.7335825 \in CI_0=[0.7044503, 0.7509661]\subset DD=[ 0.6362682, 0.9316473]$. Also, $WR_0\equiv |CI_0|/|DD| = 0.1574785 $.

The  $CI_{1}$  
is  $[0.6895103,0.7659441]$, which is also narrower than the DD interval and contains the true $B$.

\item Example 5: Consider $p=1000$ precincts all with population $N_i=150$. We let $X_i\sim Unif[0,0.7]$, $\beta_i^b\approx (N_iX_i)^{-1} Bin(N_iX_i, 1/(1+\exp(-b0-b1*X_i- (1-X_i)*\epsilon^b_i))$, $\beta_i^w \approx (N_i(1-X_i))^{-1} Bin(N_i(1-X_i),1/(1+\exp(-w0-w1*X_i- (1-X_i)\epsilon^w_i))$ (the approximation $\approx$ here involves operations such as rounding $N_iX_i$ and adding 1 to avoid zero or fractional counts), where $\epsilon^{b,w}_i$'s are $iid$ $N(0,s^2)$, $s= 1$, $b0= 0$, $b1=0$, $w0= 2.197225$, $w1=0$, $T_i\approx \beta^b_iX_i+\beta_i^w (1-X_i)$ (the approximation $\approx$ here involves operations such as replacing $X_i$ by a rounded version of $N_iX_i$ divided by $N_i$). The  resulting $T_i$ vs $X_i$ scatterplot is given by Figure \ref{fig.ex5}. 
\begin{figure}[htbp]
\includegraphics{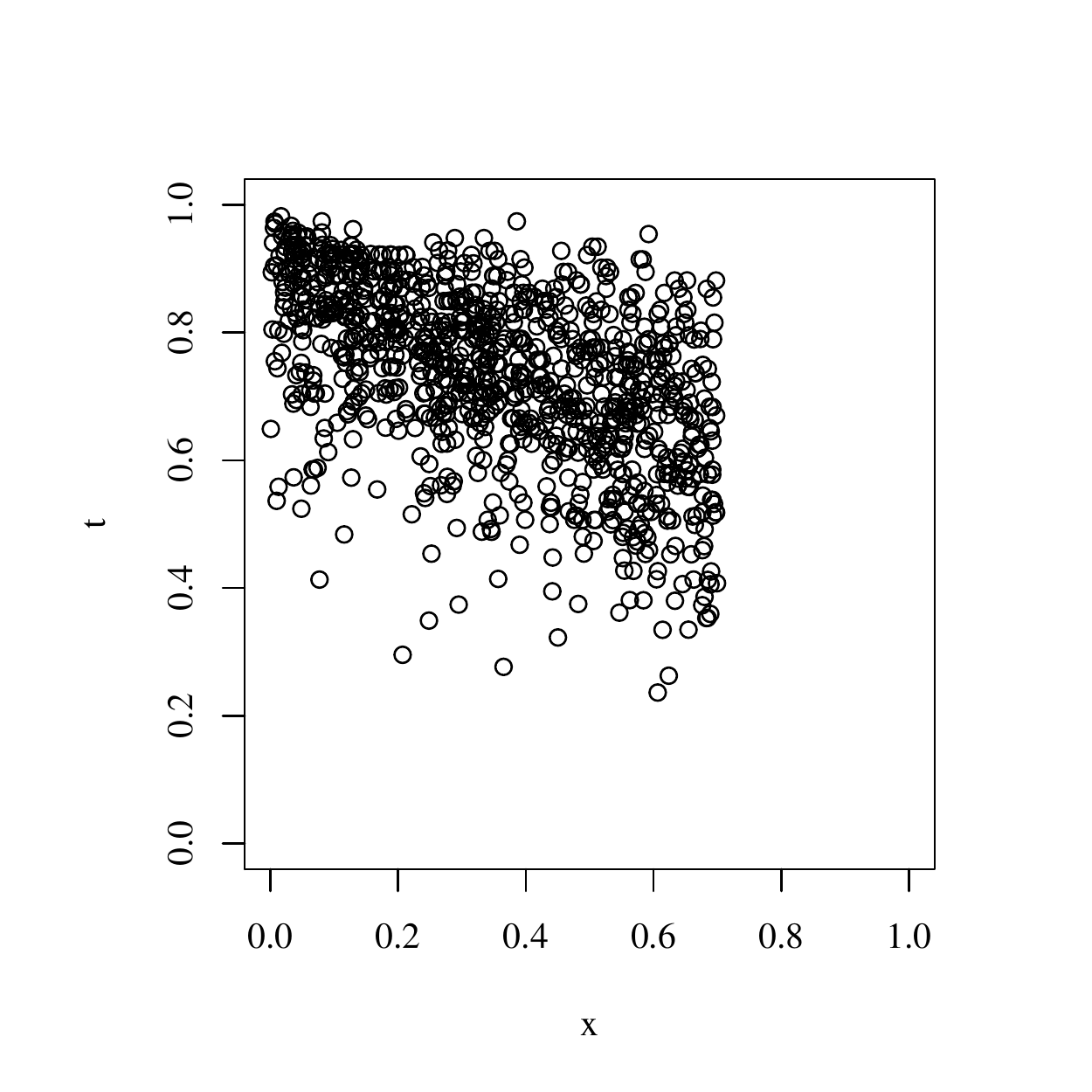}
\caption{T vs X scatterplot for Example 5.}\label{fig.ex5}
 \end{figure} 
 %
%
%
%

We apply Proposition \ref{prop2} for this example with $[l,u]=[\min(X_i),\max(X_i)]=[0.001031308,  0.6992155]$.

In this case it can be shown that $B=0.4993419 \in CI_0=[0.3998952
, 0.759834]\subset DD=[0.3403412,0.9613881]$. Also, $WR_0\equiv |CI_0|/|DD| = 0.579568 $.

The $CI_{1}$   is $[0.3711763,0.7978435]$, which is also narrower than the DD interval and contains the true $B$. 

\end{description}
 The $CI_{1}$ used in Examples 3, 4, and 5 will have at least $\Phi(1)\approx 84\%$ coverage probability asymptotically, according to Proposition \ref{propci}. In repetitions of 1000 simulations, we found that $CI_{1}$   is very conservative: $P[B\in CI_{1}]=934/1000, 1000/1000, 1000/1000$, respectively, 
 in Examples 3, 4, and 5.
The mean width of $CI_1$ divided by the mean width of the DD bound   is $0.2912078, 0.2753775,  0.6999836$, respectively.
 These results demonstrate that the proposed confidence intervals are considerably more informative about $B$ compared to the DD bounds, as shown in repeated simulations.


It is noted that in Examples 4 and 5, the true models do not follow the linear contextual model or
quadratic regression of $T_i$ vs $X_i$. The $\beta^{w,b}_i$'s follow overdispersed logistic regression model with heteroscedastic normal random effects.

%% file: sections/sec7_dataAnalyses.tex
Given that information is inherently lost in the observable information in datasets used for EI, it is important to develop models for the task on datasets with distributions similar to those used for inference. Unfortunately, for the very reason that EI is utilized in the first place, datasets with true labels in target application areas, such as elections and voting rights litigation, are typically not available for legal reasons. The nature of the learning problem is thus intrinsically different than a traditional supervised learning problem. As such, most recent work on EI has evaluated approaches using a relatively small set of datasets with ground truth on other social variables, such as voter registration and literacy, combined with artificial, simulated data (e.g., Wakefield 2004; Imai et al. 2008). Here, we significantly increase the number of datasets with ground truth labels on social data for evaluation of our proposed model, as well as to serve as a testbed for future approaches to the 2 $\times$ 2 case.

\subsection{Data}

We construct a new set of datasets for developing and evaluating EI approaches from datasets used in previous work for EI and new datasets constructed from publicly available data. Datasets from previous work (e.g., King 1997; Wakefield 2004; Imai et al. 2008) include data on voter registration and race in 1968; literacy by race in 1910; and party registration in south-east North Carolina in 2001\footnote{In the latter case, we subset the data to create 2 $\times$ 2 tables.}. We also collect data from the Centers for Disease Control and Prevention on mortality rates by gender and race (CDC 2017); literacy rates and educational attendance by gender from the 2001 Census of India (Office of the Registrar General \& Census Commissioner 2001); and additional datasets from the US Census and American Community Surveys from 1850 to 2016 via the Integrated Public Use Microdata Series (Ruggles et al. 2017). In total, we collected 459 datasets. The datasets contain a total of 2,370,854 geographic units (e.g., precincts), with an average of 5,165 geographic units per dataset and a median of 478, ranging from 145 to a maximum of 41,783. Our replication data are publicly available via Harvard Dataverse (Jiang et al. 2018).


\subsection{Analysis}
 
 Our approach aims to still produce scientifically appropriate statements (including "we don't know anything") even in the presence of (a) violations of assumptions and (b) cases where too much of the information in the individual level was aggregated away.  
Two simple heuristics are used: (I) We only consider the bounds, $CI_x$, valid if   $\hat Bl \leq \hat Bu$ and $DD$ covers part of $CI_0$; and (II) we may impose an additional restriction to only consider data with $|DD| < 0.7$. Heuristic (I) eliminates cases when the bounds flip, which can occur in practice when assumptions are violated.  (This is equivalent to applying the proposed bounds only to data sets  with nonempty $CI_0$. A theoretic support for this heuristic can be found in Appendix B, Remark \ref{remci0nonempty}.) Heuristic (II) eliminates cases in which the amount of information lost in aggregation is relatively high. Importantly, these heuristics can be applied when the ground truth is unknown, so they can be applied at inference time in real applications.
  
We observe that the proposed bounds consistently capture the true value more often than the nominal coverage intervals.  Table~\ref{table:allresults}  displays effectiveness on all of the datasets for differing levels of confidence, $\Phi(x)$, when heuristic (I) alone is applied.\footnote{As noted above, we use $\lambda=1$ in the empirical experiments.} About 63 percent of the datasets (289 out of 459) are retained after applying heuristic (I). As the confidence level increases, there is a tradeoff to be made between the capture probability of the district-level $B$ and the width-ratio.

The capture probabilities can be further improved by using both heuristics (I) and (II), see   Table~\ref{table:nohighwidthresults0.7}.
About 39 percent of the datasets (181 out of 459) are retained after applying both aforementioned heuristics.  Compared with Table~\ref{table:allresults}, p($B \in CI_x | selected)$ is increased and $E[WR_x | selected]$ is decreased when both heuristics (1) and (2) are used.
In both tables, we notice that as the confidence level increases, there is a tradeoff to be made between the capture probability of the district-level $B$ and the width-ratio. In practice, with $x=0.5$, which we consider a reasonable tradeoff between the capture probability and the width-ratio for the observed datasets, only 4 out of the 181 retained datasets are such that $B \not\in CI_x$, when both heuristics (I) and (II) are applied. The average width ratio for these 181 data sets is about 55\%, demonstrating the improvement of the proposed bound over the DD bound. The parameter of 0.7 of Heuristic (II) was chosen on a held-out, random split of the data. Increasing the value increases the proportion selected at the expense of the capture probability, and decreasing the value decreases the proportion selected to the point where datasets that are well-modeled by the approach are not selected (i.e., the problem cases tend to be those in which the DD bounds are wide). Other possible heuristics are the subject of future work, as noted in Section \ref{secgen}.

 \begin{table}[htp]
\begin{center}
\begin{tabular}{l|c|c|c|}
$x$ & $\Phi(x)$ & p($B \in CI_x | selected)$ & $E[WR_x | selected]$ \\
0.00 & 0.5000 & 0.8374 & 0.4338\\ 
0.25 & 0.5987 & 0.8927 & 0.5261\\ 
0.50 & 0.6915 & 0.9377 & 0.5977\\ 
0.75 & 0.7734 & 0.9585 & 0.6562\\ 
1.00 & 0.8413 & 0.9619 & 0.7056\\ 
1.25 & 0.8944 & 0.9654 & 0.7474\\ 
1.50 & 0.9332 & 0.9758 & 0.7856\\ 
1.75 & 0.9599 & 0.9896 & 0.8193\\ 
2.00 & 0.9772 & 0.9965 & 0.8497\\ 
\end{tabular}
\end{center}
\caption{Effectiveness in terms of the nominal coverage probability, $\Phi(x)$; proportion of intervals that capture the true district value among those selected, p($B \in CI_{x} | selected)$; and the width ratio of among those selected, $E[WR_{x} | selected]$. In this case, 62.96 percent of the datasets are selected by heuristic (I). 
} 
\label{table:allresults}
\end{table}%

\begin{table}[htp]
\begin{center}
\begin{tabular}{l|c|c|c|}
$x$ & $\Phi(x)$ & p($B \in CI_x | selected)$ & $E[WR_x | selected]$ \\
0.00 & 0.5000 & 0.8564 & 0.3652\\ 
0.25 & 0.5987 & 0.9227 & 0.4718\\ 
0.50 & 0.6915 & 0.9779 & 0.5502\\ 
0.75 & 0.7734 & 0.9945 & 0.6120\\ 
1.00 & 0.8413 & 0.9945 & 0.6640\\ 
1.25 & 0.8944 & 0.9945 & 0.7091\\ 
1.50 & 0.9332 & 0.9945 & 0.7519\\ 
1.75 & 0.9599 & 1.0000 & 0.7904\\ 
2.00 & 0.9772 & 1.0000 & 0.8261\\ 
\end{tabular}
\end{center}
\caption{Effectiveness in terms of the nominal coverage probability, $\Phi(x)$; proportion of intervals that capture the true district value among those selected, p($B \in CI_x | selected)$; and the width ratio of among those selected, $E[WR_x | selected]$. In this case, 39.43 percent of the datasets are selected by heuristics (I) and (II).
}
\label{table:nohighwidthresults0.7}
\end{table}

%% file: sections/sec8_limitations.tex
Our work  makes a single pair of essential assumptions (\ref{asw}) and (\ref{asb}) on the linear contextual effects. This is  more general than the traditional methods which assume zero contextual effects, which are often falsified by real data with knowledge of the ground truth.

When assumptions 
 (\ref{asw}) and (\ref{asb}) fail, it can be shown that our method does not always work. The key question in practice is how can one know about such violations. Unfortunately, it is theoretically possible that
such violations can not be detected by data without knowledge of the ground truth. Consider the following example:

\begin{description}
\item Example 6:  Suppose  $X_i \sim Unif[0,1]$ and $N_i$ is independent of $X_i$ and $\beta^{b,w}_i$. Suppose we have quadratic contextual effects
$\beta^b_i=T+b_2 (X^2_i-1)$, $\beta^w=T+b_2 (X_i^2+X_i)$,
where to ensure these are probabilities valued in $[0,1]$ for all possible $X$, we restrict $T\in(0,1)$ and $b_2\in [\max\{-T/2, -(1-T)\}, \min\{T,(1-T)/2\}]$. Then $T_i=\beta^b_iX_i+\beta^w_i(1-X_i)=T$. The observed data $(X_i,T_i)$ would be the same as  our Example 1 earlier ($T_i=T$). We have already found that the large sample limit of our proposed bound is $CI_0=T\pm (1/3)\min\{T,1-T\}$. The large sample limit of true $B$ is now $E(N_iX_i\beta_i^b)/E(N_iX_i)=T-b_2/2$. It is then possible
that for large enough $b_2$,  $B\not\in CI_0$ (e.g, when $b_2=T=1/3$).  The same holds  in the large sample limit for  $CI_x$ with any $x>0$,  since the sampling variation that differentiates between $CI_x$ and $CI_0$ disappears  in the large sample limit. 
\end{description}

If all datasets were generated from this model (e.g., with $b_2=T=1/3$), then the asymptotic coverage probability of any $CI_x$ would be 0 and we would not be able to avoid such data sets without the knowledge of the ground truth. Fortunately, this kind of ``non-detectable violation'' happens quite rarely.  For example, the non-detectable violation  in Example 6  is caused by the  quadratic effects in $\beta^b_i$ and $\beta^w_i$ canceling each other exactly by chance. In addition, our interval estimates are robust in the sense that even a small amount of violation of the assumptions would not matter. For example, the quadratic effect $b_2$ does not have to be exactly 0 for $CI_0$ to capture $B$.  This is in contrast to traditional point estimates and their confidence intervals, which will miss the true parameter due to any bias   when the sample size $p$ is sufficiently large, since the width of the confidence interval typically shrinks at the rate of $1/\sqrt{p}$.


From hundreds of real data sets on which we evaluated the approach, we found that most practically important violations  can be easily detected if $CI_0$ is empty   (i.e.,  the regression bound either flips, or  does not intersect with the DD bound at all).     Appendix B examines   this analytically  for the limit of large $p$ (see Remark \ref{remci0nonempty}). 
 The logic there is to prove that if the assumptions hold, then $CI_0$ should not be empty. Therefore if  
$CI_0$ is found to be empty, then something must be wrong about the assumptions.

As shown in Section \ref{secreal}, we tried hundreds of real data sets (with knowledge of the ground truth), and in most cases, we have nonempty $CI_0$.  When applying the $CI_x$ for $x>0$ on the selected data sets with nonempty $CI_0$,  we found  that   our conservative confidence interval $CI_x$ tends to capture the  true district parameter $B$ more often than the  stated level of confidence $\Phi(x)$, while tightening the DD bound. For example, 
$CI_{0.5}$ has nominal coverage probability about $70\%$, but it actually captures $B$ more than $90\%$ of the selected data sets.\footnote{As Section \ref{secreal} indicates, the actual percentage of captures may be further improved with an additional restriction on studying data sets with $|DD|<0.7$.} For  the selected data sets where $CI_{0.5}$ misses $B$, most of them are data sets in which the $X_i$'s represent the gender proportion in a particular precinct. Gender data are known to be problematic for EI. Their $X_i$'s tend to focus on a short range near $0.5$, with possibly influential outliers near $0$ or $1$.  

In addition to data with a short range of $X$, data with influential observations may also cause the proposed method to fail (either by not selecting the data via the heuristics, or by selecting the data and missing the true district level value) when  data points  $(X_i, T_i)$ seem to belong to several different clusters. In these situations, we found that a divide and conquer strategy may be helpful.  One could divide the data into several parts and apply either the proposed bound or the DD bound to each part, depending on the observed pattern in the particular part. The proposed bound could be applied to any part of the data that displays a common pattern (e.g., those of linear or quadratic regression). For parts of the data that are outliers or that otherwise lack a clear pattern for linear or quadratic regression, one could apply the DD bound. The bounds would then be combined by weighting the number of people in each part of the data to obtain a single bound.
In initial experiments of such an approach, we segmented the data visually and found that this strategy can sometimes rectify the misses or nonselection of the current method. We leave automating the process of segmentation to future work.

%% file: sections/sec9_discussion.tex
 An alternative approach is to assume nonlinear contextual effects such as $E(\beta_i^b|X_i)=1/(1+e^{-b_0-b_1X_i})$ and  $E(\beta_i^w|X_i)=1/(1+e^{-w_0-w_1X_i})$. At first sight this seems to avoid the non-identifabililty problem in the model $E(T_i|X_i)=X_i/(1+e^{-b_0-b_1X_i})
 + (1-X_i)/(1+e^{-w_0-w_1X_i})$. 
  However, the limitations of such an approach are noted by
 Wakefield (2004, Section 1.3): 
 \begin{quote}
 Unfortunately, assuming nonlinearity theoretically
removes the nonidentifiability but in practice is totally dependent on the form chosen, and
parameter estimates will in general be highly unstable. This was pointed out by Achen and Shively (1995: 117), who comment that since the contextual effects are not strong and the
range of $X$ is often not (0, 1), it would be virtually impossible to discriminate between nonlinear and linear forms (since any function that has a narrow range and does not change greatly may be well approximated with a linear form, via a Taylor series expansion).
\end{quote}
In contrast, our current work directly confronts the non-identifiability problem by modeling the linear contextual effects and generates interesting insights on the bounds of the unidentified parameter and on the sensitivity of its effect, which can not be easily derived in the nonlinear model where unstableness of the point estimates is hidden in a nontrivial way.

Our model has derived how the district level parameter depends on the non-identified parameter on linear contextual effect. There is only one such non-identified parameter, which could allow sensitivity analysis, such as based on (\ref{eqsensitive}).

  The current work only considers the black proportion $X_i$ for the contextual effect. One may consider adding other covariates to the contextual effect models
for modeling $\beta^b_i$ and $\beta^w_i$.
 The current paper only focuses on inference regarding the district level parameter. It would be interesting to obtain a useful bound for the precinct level parameters $\beta^b_i$,
 probably by modeling the distribution of the residuals $(\beta^b_i-E(\beta^b_i|X_i), (\beta^w_i,E(\beta^w_i|X_i))$, or at least the second moments such as $var((\beta^b_i,\beta^w_i)^T |X_i )$. (The residuals average out in the district level estimates,  so we could still get useful bounds for the district level parameter in the current paper, even without modeling the residuals.)
Also, it would be interesting to extend the idea of this paper to the case of more general RxC tables,  for which   Cho and Manski (2008) have derived model-free bounds  that generalize those of Duncan and Davis (1953).

%% file: sections/sec_appendix_a.tex
Note that from (\ref{bexp}),  $\beta^b_i=b_i(\lambda,w_1,\theta)+(e^b_i-\lambda e^T_i)$,
 where the residual $(e^b_i-\lambda e^T_i)$ has mean 0.  
 
 For the district level parameter, the residuals can be averaged out  over many precincts  due to the central limit theorem and we can get a  potentially useful conservative confidence interval, without  modeling the variance of the residuals:
$$
 B=\sum_iN_iX_i\beta^b_i/\sum_iN_iX_i=\sum_iN_iX_i[b_i(\lambda,w_1,\theta)+(e^b_i-\lambda e^T_i)] /\sum_iN_iX_i$$
 $$
 B= \frac{\sum_iN_iX_i(e^b_i-\lambda e^T_i)}{ \sum_iN_iX_i} +\frac{\sum_iN_iX_ib_i(\lambda,0,\theta) }{\sum_iN_iX_i}-w_1\frac{\sum_iN_iX_i (1-X_i)}{\sum_iN_iX_i}.
 $$
The unidentified parameter $w_1\in[wl,wu]$.

Therefore $B\in[BL(\theta),BU(\theta)]$, where
$$BL(\theta)\equiv \frac{\sum_iN_iX_i(e^b_i-\lambda e^T_i)}{ \sum_iN_iX_i} +\frac{\sum_iN_iX_ib_i(\lambda,0,\theta) }{\sum_iN_iX_i}-wu(\theta)\frac{\sum_iN_iX_i (1-X_i)}{\sum_iN_iX_i};
$$
$$BU(\theta)\equiv
\frac{\sum_iN_iX_i(e^b_i-\lambda e^T_i)}{ \sum_iN_iX_i} +\frac{\sum_iN_iX_ib_i(\lambda,0,\theta) }{\sum_iN_iX_i}-wl(\theta) \frac{\sum_iN_iX_i (1-X_i)}{\sum_iN_iX_i}.
$$

Here $wl,wu$ depend linearly on $\theta\equiv (w_0,c_1 ,d_1 )$.
 The $b_i(\lambda,0,\theta)\equiv b^0_i+(b^1_i)^T\theta \equiv \lambda T_i +
(1-\lambda, 1-\lambda X_i, X_i-\lambda X_i^2) ( w_0,c_1,d_1)^T$
  also depends linearly on $\theta\equiv (w_0,c_1 ,d_1 )$,   which is estimated by quadratic
 regression (\ref{regr}) as $\hat \theta\equiv (\hat w_0,\hat c_1, \hat d_1 )$, with robust asymptotic variance matrix $V=\hat avar(\hat\theta)$ based on a sandwich formula.\footnote{See, e.g., https://www.stata.com/manuals/p\_robust.pdf}

 Denote the first term in $BL(\theta)$ or $BU(\theta)$ as
$$
TERM_1=\frac{\sum_iN_iX_i(e^b_i-\lambda e^T_i)}{ \sum_iN_iX_i}.
$$
Then $E(TERM_1)=0$. 
Assuming independent precincts, then the first term $TERM_1$  has asymptotic variance  $Var(TERM_1)=\sum_i\left(\frac{N_iX_i   }{ \sum_iN_iX_i}\right)^2var(e^b_i-\lambda e^T_i|N_i,X_i)$. Note that $var(e^b_i-\lambda e^T_i|N_i,X_i) =var(\beta^b_i-\lambda T_i|N_i,X_i)
=var((1-\lambda X_i)\beta^b_i-\lambda(1-X_i)\beta^w_i)|N_i,X_i)\equiv var(\epsilon_i|N_i,X_i)$,
where $\beta^b_i$ and $\beta^w_i$ are probabilities valued in $[0,1]$. 
Then $\epsilon\equiv (1-\lambda X_i)\beta^b_i-\lambda(1-X_i)\beta^w_i)$
has a range $[-\lambda(1-X_i), 1-\lambda X_i]$, if we restrict $\lambda\in[0,1/X_i]$.   The variance of a bounded random variable in $[a,b]$ is at most $[(b-a)/2]^2$. Therefore, $var(\epsilon|N_i,X_i)\leq [(1+\lambda)/2-\lambda X_i]^2 
$ and $Var(TERM_1)\leq \sum_i\left(\frac{N_iX_i  [(1+\lambda)/2-\lambda X_i] }{ \sum_iN_iX_i}\right)^2 $.
Therefore, we know that the asymptotic standard error of the first term is bounded above by
$$sd(TERM_1)\leq S_1= \sqrt{  \sum_{i=1}^p\left(\frac{N_iX_i[(1+\lambda)/2-\lambda X_i]    }{ \sum_{i=1}^p N_iX_i}\right)^2 },
 $$ 
if we choose $\lambda\in[0,1]$ (which guarantees $\lambda\in[0,1/X_i]$ for all $i$).

Now $wl=wl(\theta)$ is of the form $\max_{j=1}^J\{gl^0_j+gl_j^T\theta\}$
for some constant vectors $gl_{j}$;  $wu=wu(\theta)$ is of the form $\min_{j=1}^J\{gu^0_j+gu_j^T\theta \}$, 
for some constant vectors $gu_{j}$.  Denote 
$r\equiv \frac{\sum_iN_iX_i (1-X_i)}{\sum_iN_iX_i}$, $h_0\equiv\frac{\sum_iN_iX_i b^0_i }{\sum_iN_iX_i}$,
$h\equiv\frac{\sum_iN_iX_i b^1_i  }{\sum_iN_iX_i}$.

Then
$$BL(\theta)=TERM_1 +\frac{\sum_iN_iX_i(b^0_i+(b^1_i)^T\theta) }{\sum_iN_iX_i}-\min_{j=1}^J\{gu^0_j+gu_j^T\theta \} r$$$$
=TERM_1+h_0+h^T\theta -\min_{j=1}^J\{gu^0_j+gu_j^T\theta \} r.$$

We can write 
$BL(\theta)=\max_{j=1}^J\{BL_j\}$
where
$BL_{j}=TERM_1+h_0-rgu^0_{j}+(h- rgu_{j})^T\theta$.
Similarly, we can write 
$BU(\theta)=\min_{j=1}^J\{BU_j \}$
where
$BU_{j}=TERM_1+h_0-rgl^0_{j}+(h- rgl_{j})^T\theta$.
    
Now define
\begin{equation}
\hat BL=\max_{j=1}^J\{\hat BL_j \},\end{equation}
 where
$\hat BL_{j}= h_0-rgu^0_{j}+(h- rgu_{j})^T\hat \theta$;
\begin{equation}
\hat BU=\min_{j=1}^J\{\hat BU_j \}, \end{equation}
 where
$\hat BU_{j}= h_0-rgl^0_{j}+(h- rgl_{j})^T\hat \theta$.

 [It can be verified that in the previous notation of (\ref{intest}), we   have $\hat BL=B(\lambda,wl(\hat\theta),\hat\theta)$ 
and $\hat BU=B(\lambda,wu(\hat\theta),\hat\theta)$. ] 

Note that $$\hat BL_{j}-BL_{j}=-TERM_1 +(h- rgu_{j})^T(\hat\theta-\theta);$$ 
$$\hat BU_{j}-BU_{j}=-TERM_1 +(h- rgl_{j})^T(\hat\theta-\theta).$$ 
By an asymptotic normality argument,
$\hat BL_{j} \approx N(BL_{j}, sl_{j}^2)$
where $sl_{j}\leq SL_{j}\equiv S_1+\sqrt{(h-rgu_{j})^TV(h-rgu_{j})}$;
$\hat BU_{j}  \approx N(BU_{j}, su_{j}^2)$
where $su_{j}\leq SU_{j}\equiv S_1+\sqrt{(h-rgl_{j})^TV(h-rgl_{j})}$, for all $j=1,...,J$.  Assuming $V$ is of order $O_{\rm p}(1/ p )$, then all $SU_j$ and $SL_j$'s are also of order $ O_{\rm p}(1/\sqrt{p})$. The sample variations
$\hat BU_{j}- BU_{j}$ and $ \hat BL_{j}- BL_{j}$ are also of order $ O_{\rm p}(1/\sqrt{p})$.

Now consider various cases of the bound $B\in[BL(\theta),BL(\theta)]$.
 Assume that  $N_iX_i(1-X_i)$ is not almost surely 0, then  the large sample limit of the sensitivity parameter $  \frac{BU(\theta)-BL(\theta) }{wu(\theta)-wl(\theta)} =r=  \frac{\sum_iN_iX_i (1-X_i)}{\sum_iN_iX_i}$ is a positive number due to the law of large numbers.  Assume that 
$wu(\theta)\neq wl(\theta)$ (and therefore $BU(\theta)\neq BL(\theta)$). Then $w_1$ can be $close$ (within $O_{\rm p}(1/\sqrt{p})$) to only one of the end points of $[wl(\theta),wu(\theta)]$, and consequently  $B$ can be close to only one end point of $[BL(\theta),BU(\theta)]$. Without loss of generality we assume that $B$ is close to $BL(\theta)$.   (The other possibility would be similar.) 
Assume that the  minimizing entry of $wu= \min_{j=1}^J\{gu^0_j+gu_j^T\theta \}$ is  unique and not tied with the other entries. Then the maximizing entry   $BL(\theta)=max_{j=1}^J\{BL_j\}$ is unique and has an order-1 gap from the other entries, that is greater than $ O_{\rm p}(1/\sqrt{p})$, which is the order of all $(\hat BU_{j}- BU_{j})$'s and $ (\hat BL_{j}- BL_{j})$'s. Therefore
$\max_{j=1}^J \{\hat BL_{j}\}$ ($=\hat BL$) and  $\max_{j=1}^J\{BL_j\}$ ($= BL(\theta)$) are achieved at a same $j$, with probability tending to 1 as $p\rightarrow \infty$. We will call this same $j$ as $\hat j$.    
Then 
$$
\hat BL=\hat BL_{\hat j}=BL_{\hat j} + (\hat BL_{\hat j}-  BL_{\hat j})
$$
$$
=BL(\theta) + (\hat BL_{\hat j}-  BL_{\hat j})
$$
where the last term is asymptotically normal and of order $O_{\rm p}(1/\sqrt{p})$.
 
There is a similar equation relating $\hat BU$ to $BU(\theta)$. These imply that the $\hat BU$ is close to $BU(\theta)$.

Since we have assumed that $B$ is only close to one end point $BL(\theta)$, and not close to $BU(\theta)$, then $B$  must not  be close to  $\hat BU$ or $\hat BU+u'$ either, for any $u'$ of order $O_{\rm p}(1/\sqrt{p})$.
 Then we have
$P(B>\hat BU+u')$ converges to 0. Then
$P(B\not\in [\hat BL-l',\hat BU+u']) \approx  P(B<\hat BL-l') \approx
P(B<  BL(\theta) -l' + (\hat BL_{\hat j}-  BL_{\hat j}))$
$\leq P( l'  < (\hat BL_{\hat j}-  BL_{\hat j}))$
since $B>  BL(\theta)$. 
Then $  P( l'  < (\hat BL_{\hat j}-  BL_{\hat j}))$
$\approx \Phi (-l'/sl_{\hat j}) \leq \Phi (-l'/SL_{\hat j})=\Phi(-x)$
if we take $l'=x SL$. Setting $u'=xSU$ of order $O_{\rm p}(1/\sqrt{p})$ leads to an approximate  upper bound of $P(B\not\in [\hat BL-l',\hat BU+u'])$  being $1-\Phi(-x)=\Phi(x)$ (for large $p$).
  Q.E.D.

\begin{rem} 
  The coverage probability of $CI_x$ can be improved to 1,  if  
we have $w_1$ lying in the interior of the bound $(wl,wu)$. This would allow any $x>0$ to be used in finding a confidence interval. However, the condition on $w_1$ cannot be checked due to its non-identifiability.   The tie-breaking conditions that we assumed  about the identified $\theta$, however, can be checked by data. Then we can, e.g.,  use $x=1$ and achieve coverage probability at least $ \Phi( x)\approx 84\%$, or use $x=1.282$ and achieve at least $90\%$ coverage probability.
 \end{rem}

%% file: sections/sec_appendix_b.tex
In this Appendix B, we prove in the large $p$ limit that when model assumptions hold, $CI_0$ should be nonempty.  

 By tracing the definition of $CI_0$ and applying the Law of Large Numbers,
we find that in the large sample limit  $$CI_0=[\inf_{v_1\in  [wl_j,wu_j]}B(\lambda,v_1,\theta),\sup_{v_1\in  [wl_j,wu_j]}B(\lambda,v_1,\theta)]\cap DD,$$ where
$DD$ denotes the larges sample limit of the DD bound,
$$B(\lambda,v_1,\theta)=E\{N_iX_i[(w_0+c_1+d_1X_i)+v_1(X_i-1)]\}/E\{N_iX_i\}$$ 
as summarized in (\ref{Bformula}) and (\ref{bformula}) before, and
$[wl_j,wu_j], \; j\in\{1,2,3\}$ indicate the bound of $w_1$ to be used according to the $j$th Proposition.
\begin{prop} (Non-emptiness of $CI_0$.) For $j\in\{1,2,3\}$,  assume linear contextual effects $E(\beta_i^w|X_i,N_i)=w_0+w_1X_i$ and  $E(\beta_i^w|X_i,N_i)=b_0+b_1X_i$, and let $[wl_j,wu_j] $ indicate the bound of $w_1$ to be used according to the $j$th Proposition.
  Define   in the large sample limit  
$$CI_0=[\inf_{v_1\in  [wl_j,wu_j]}B(\lambda,v_1,\theta),\sup_{v_1\in  [wl_j,wu_j]}B(\lambda,v_1,\theta)]\cap DD,$$ where 
$$DD=[E[ N_i \max\{ 0, T_i-(1-X_i)\} ] / E( N_iX_i),      E[N_i \min\{ T_i, X_i\}  ]/E(N_iX_i)     ],$$ and
$$B(\lambda,v_1,\theta)=E\{N_iX_i[(w_0+c_1+d_1X_i)+v_1(X_i-1)]\}/E\{N_iX_i\},$$  where $c_1,d_1$ follow  (\ref{cdeq}).

Then $CI_0$ is nonempty.
\end{prop}

Proof:

 $$B(\lambda,v_1,\theta)=E\{N_iX_i[(w_0+c_1+d_1X_i)+v_1(X_i-1)]\}/E\{N_iX_i\}$$ 
$$=E \{N_iX_i[(w_0+c_1X_i+d_1X_i^2 -(w_0+v_1X_i)(1-X_i)]/X_i\}/E\{N_iX_i\}
$$ 
$$=E \{N_iX_i[(T_i -(w_0+v_1X_i)(1-X_i)]/X_i\}/E\{N_iX_i\} 
$$
$$=E \{N_iX_i[ \beta^b_iX_i+\beta^w_i(1-X_i) -(w_0+v_1X_i)(1-X_i)]/X_i\}/E\{N_iX_i\} 
$$
$$=E \{N_iX_i[ \beta^b_iX_i+(w_0+w_1X_i)(1-X_i) -(w_0+v_1X_i)(1-X_i)]/X_i\}/E\{N_iX_i\} 
$$
$$=E \{N_iX_i[ \beta^b_i +( w_1-v_1) (1-X_i) \}/E\{N_iX_i\} 
$$
$$\equiv B+( w_1-v_1) r  . 
$$
where we denote $r=E \{N_iX_i  (1-X_i) \}/E\{N_iX_i\}$ and $B=E \{N_iX_i \beta^b_i  \}/E\{N_iX_i\}$.
Then
$$CI_0=[B+(w_1-wu_j)r,  B+(w_1-wl_j)r ]\cap DD.$$

On the other hand, the $j$th Proposition implies that $$w_1\in [wl_j,wu_j] . $$ Then $$B\in [B+(w_1-wu_j)r,  B+(w_1-wl_j)r ],$$ since $r\geq 0$.
Now we also have  $$B\in DD,$$ since $$X_i \beta^b_i =T_i-(1-X_i) \beta^w_i \in  [\max\{ 0, T_i-(1-X_i)\}, \min\{ T_i, X_i\}]$$ due to Duncan and Davis (1953). Then we have  $$B\in[B+(w_1-wu_j)r,  B+(w_1-wl_j)r ]\cap DD =CI_0$$ in the large sample limit.
Therefore $CI_0$ is non-empty.

Q.E.D.

\begin{rem}\label{remci0nonempty}
 In practice,  one can apply the converse of this Proposition to  rule out data sets with empty $CI_0$, which likely suggests either some assumptions are violated or the size of the data is not large enough for the method to work reliably.   The logic is that the interval should not be empty if the assumptions all hold and the sample size is large enough.
\end{rem}